\documentclass[mathleft]{an}
\usepackage{graphicx,amsmath,amssymb}
\graphicspath{{/home/ap1plc/articles/figures/}}
\usepackage{times}
\def\f#1#2{{#1 \over #2}}

\newcommand{\lam}{\lambda}

\newcommand{\be}{\beta}

\newcommand{\al}{\alpha}

\newcommand{\sig}{\sigma}

\newcommand{\Om}{\Omega}
\newcommand{\om}{\omega}

\begin{document}

\Pagespan{1}{}
\Yearpublication{2007}%
\Yearsubmission{2007}%
\Month{11}%
\Volume{999}%
\Issue{88}%

\title{Absolute and convective instabilities in an inviscid compressible
mixing layer}

\author{P. Caillol\thanks{Corresponding author:
  \email{p.l.caillol@sheffield.ac.uk}\newline}
\and  M. Ruderman }
\titlerunning{Instabilities in a compressible
mixing layer}
\authorrunning{P. Caillol \& M. Ruderman}
\institute{
Department of Applied Mathematics, Sheffield University,
Hicks Building, Hounsfield Road, Sheffield S3 7RH, UK}

\received{30 April 2007}
\accepted{}
\publonline{later}


\abstract{
 We consider the stability of a compressible shear flow separating
two streams of different speeds and temperatures. The velocity and
temperature profiles in this mixing layer are hyperbolic tangents. 
 The normal mode analysis of the flow stability reduces to an eigenvalue 
problem for the pressure perturbation. 
We briefly describe the numerical method that we
used to solve this problem. Then, we introduce the notions of the absolute and
convective instabilities and
examine the effects of Mach number, and 
the velocity and temperature ratios of each stream
on the transition between convective and absolute instabilities.
Finally, we discuss the implication of the results presented in
this paper for the heliopause stability.}

\maketitle

\section{Introduction}
\label{sec:intro}

The theory of stability of shear flows has been the subject of an intense study
for the past century due to its numerous applications in engineering,
geophysics, and astrophysics. Velocity shear generates a dynamical instability,
and the most common and fastest one is the Kelvin-Helmholtz (KH) instability
which occurs in fluid and plasma inviscid shear layers. It intervenes in
many phenomena in fluid dynamics (Blumen et al. 1975; Jackson \&
Grosch 1989; Caillol 2005), in space, astrophysical and laboratory plasmas  
(Miura 1982; Jones et al. 1997; Mills et al. 2000; Ruderman 2000;
Taro\-yan \& Erd\'elyi 2002, Hasegawa 2004) when
steep velocity gradients emerge. The most relevant examples are the
stability of the interface between the solar wind and the planetary
magnetospheres, the interaction between adjacent streams of different
velocities in the solar wind and polar cusps, the dynamic structure of
cometary tails, and the interaction between the solar wind and the interstellar
medium.

In space and astrophysical plasmas, the total velocity jump across the velocity
shear layer can have large sonic Mach number, so that taking into account the
compressibility of plasmas is essential. KH instability weakens as the
convective Mach number increases (Mach number in the frame moving with a
perturbation), while oblique disturbances become more and more unstable and
dominant.

In astrophysical applications, the account of magnetic field is very often
important. In that case, the fluid motion is described by the
magnetohydrodynamic (MHD) rather than hydrodynamic (HD) equations. Studying
the MHD KH instability is much more complicated than the HD one due to the
coupling between KH instabilities and Alfv\'en field line resonances
(Hollweg et al. 1990; Yang \& Hollweg 1991; Ruderman \& Wright 1998;
Taroyan \& Erd\'elyi 2003). In this paper, we study the HD KH instability
postponing the magnetic field account for future work.

The study of stability of compressible hydrodynamic shear flows has a long
history. Landau (1944) analytically studied the discontinuous flow
generated by a vortex sheet and proved its stability when the
Mach number $M$ is greater than $\sqrt{2}$ with
respect to two-dimensional disturbances.
Blumen et al. (1975) showed that
the inviscid flow with the velocity profile $U(y)=\tanh(y)$, where $y$ is the
Cartesian coordinate perpendicular to the flow direction, is temporally
unstable for any $M$.

The normal mode analysis used in these studies revealed many important
properties of the KH instability of shear flows. However, it failed to predict
if a finite portion of flow looks stable or unstable in a fixed reference
frame. The reason is that the normal mode analysis deals with periodic
perturbations infinite in at least one spatial direction, while real
perturbations always occupy a finite spatial domain. To answer this question,
we need to solve the initial value problem for an arbitrary perturbation
bounded in a finite domain. Then two scenarios are possible. In the first
scenario, the perturbation exponentially grows with time at any fixed spatial
position. This
type of instability is called {\em absolute}\/. In the second scenario, the
perturbation exponentially grows with time, but in the same time it is
convected
out of the observational domain so fast that eventually it decays at any fixed
spatial position. This type of instability is called {\em convective}\/.

The notions of the absolute and convective instabilities were first introduced
in plasma physics (see Briggs 1964; Bers 1973). Later, it started to be used in
hydrodynamics (e.g., Kulikovskii \& Shikina 1977; Huerre \& Monkewitz 1985;
Brevdo 1988). Initially, the absolute and convective instabilities were studied
for shear flows of incompressible fluids. Pavithran \& Redekopp (1989) analysed
the convective and absolute instabilities in a shear flow of a compressible
fluid when the velocity and temperature have hyperbo\-lic-tangent profiles.
Jackson \& Grosch (1990) carried out a similar study for a boundary layer flow
where the velocity field is described by the hyperbolic tangent, and the
temperature is linked to the velocity by the so-called Crocco relation.
In both papers, the analysis is restricted to the subsonic regime.
Terra-Homen \& Erd\'elyi (2003) determined the transition from the convective
to the absolute instability in a shear flow of a viscous compressible fluid by
solving the initial value problem for the full nonlinear system of hydrodynamic
equations. In their study, they considered a shear flow with a constant
temperature and the hyperbolic-tangent velocity profile.
They also restricted their
analysis to subsonic flows, and took the Reynold number equal to $10^3$. With
these constraints, the stabilizing effect of viscosity was stronger than that
of compressibility.

The aim of this paper is to study the transition from the convective to the
absolute instability in the same shear flow as one considered by Pavithran
\& Redekopp (1989), however with the Mach number attaining supersonic values.
The paper is organized as follows. In the next section, we formulate the
problem, and derive the governing equation and the boundary conditions for the
pressure perturbation. In Sect.~\ref{sec:spectr}, we describe the spectral
method used to solve the eigenvalue problem determining the flow stability with
respect to normal modes.
In Sect.~\ref{sec:convect}, we present the results of the study of the flow
absolute and convective instabilities. In Sect.~\ref{sec:helio}, we discuss the
implication of the results obtained in this paper for the heliopause stability.
Sect.~\ref{sec:concl} displays the summary of the results and our
conclusions.
  
\section{Formulation}
\label{sec:formul}

\begin{figure}
\includegraphics[width=80mm,height=50mm]{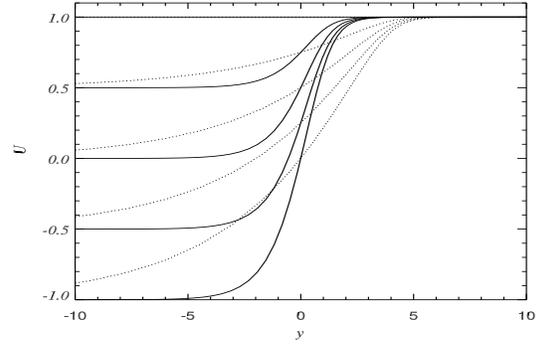}
\vspace*{-2mm}
\caption{Profile $U(y)$ for various values of $\be_U$, $\be_T=2$ (solid line)
and $\be_T=10$ (dotted line).}
\label{profil1}
\end{figure}

\begin{figure}
\includegraphics[width=80mm,height=50mm]{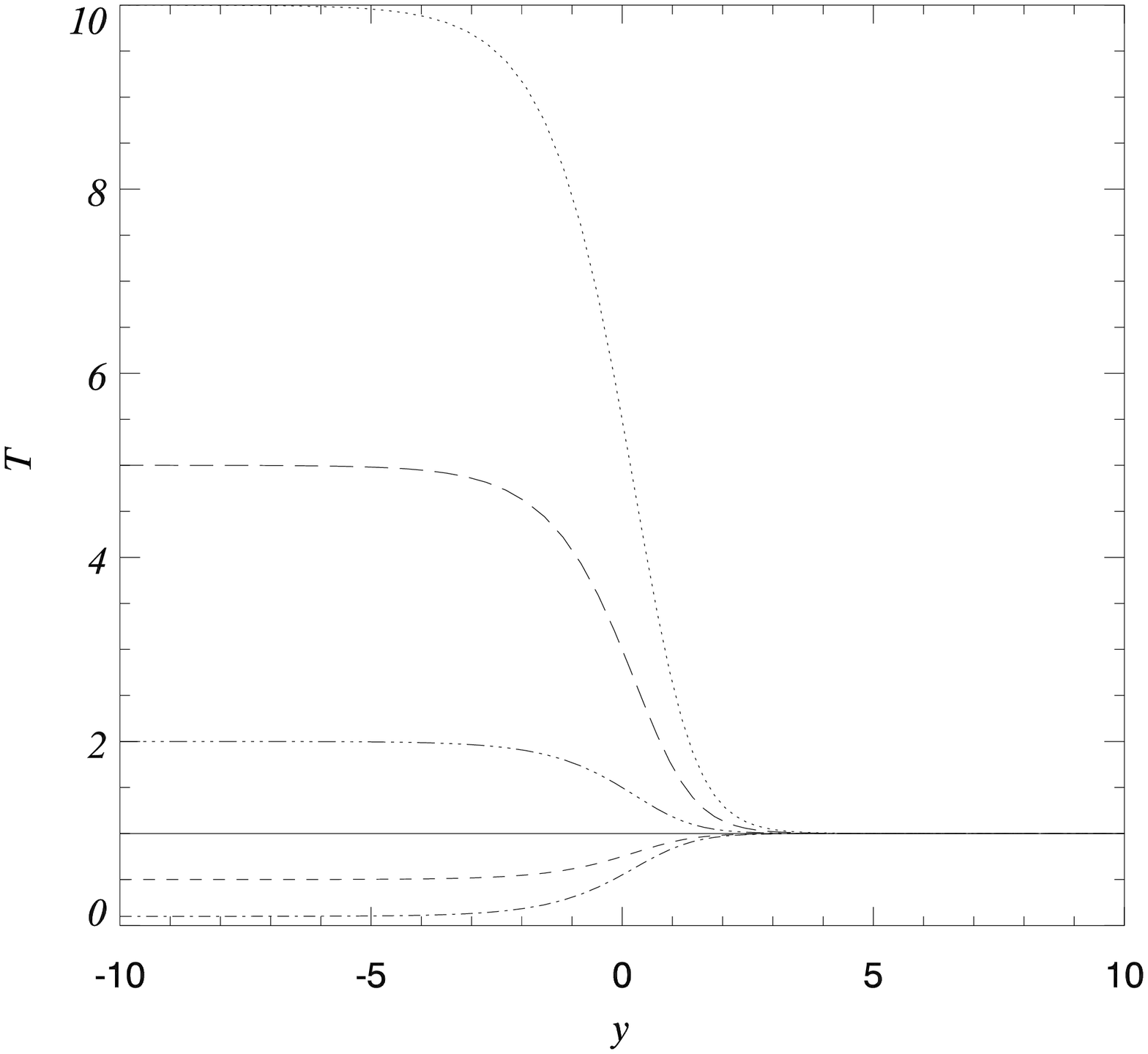}
\vspace*{-2mm}
\caption{Profile $T(y)$ for $\be_T = 0.1,\ 0.5,\ 1,\ 2,\ 5$ and $10$.}
\label{profil2}
\end{figure}

We consider the instability of a two-dimensional and unbounded 
compressible ideal shear flow. The current is in the
$x$\/-direction. Both velocity and temperature magnitudes are functions
of $y$ in Cartesian coordinates $x,y,z$\/  
and are given by
\begin{eqnarray}
U &=& \mbox{$\f12$} [1+\be_U +(1-\be_U) \tanh (\eta)]\,,\\
T &=& \mbox{$\f12$} [1+\be_T +(1-\be_T) \tanh (\eta)]\,,\label{tlay}
\end{eqnarray}
where $\eta$ is the cross-stream variable in the Howarth-Dorod\-nitzyn
transformation, $\,\eta=\int_0^y 1/T(y')\,dy'\:$ (Jackson \& Grosch 1989).
The velocity and temperature are made dimensionless using their values in the
fast stream ($y \to \infty$). The spatial variables and time are also
dimensionless and measured in units of the half-width of the shear layer 
$y_0$\/, and $y_0$ divided by the fast stream
velocity, respectively. The quantity $\be_U$ is the ratio of
the speed of the slow stream to that of the fast stream, and $\be_T$ is the same
but for temperature. We always have
$$-1 <\be_U < 1\,,\qquad \be_T >0,$$
$$U \to \be_U, \quad T\to \be_T \quad \mbox{at} \;\;\eta\to -\infty.$$
The velocity and temperature profiles are shown for different values of $\be_U$
and $\be_T$ in Fig.~\ref{profil1} and Fig.~\ref{profil2}
respectively. Flows of this type represent transition layers between two
streams with constant parameters. The KH instability results in mixing of
the fluids from both streams inside the transition layer.
This is why such flows are called mixing layers.

The flow field is perturbed by introducing wave dis\-tur\-ban\-ces whose
amplitudes are functions of $\eta$; for example, the pressure perturbation is
$$p =\Pi(\eta)\,\exp[i(\hat{\al} x +\hat{\be} z -\hat{\om} t)]\,,$$
with $\hat{\al}$ and $\hat{\be}$ the wavenumbers and $\hat{\om}$ the frequency.
The equation governing $\Pi$ (Jackson \& Grosh 1989) is
\begin{eqnarray}\label{comp1}
\Pi'' -\frac{2 U'\Pi'}{U-\hat{c}}
&-& T\, [(\hat{\al}^2+\hat{\be}^2)T \nonumber\\
&-& \hat{\al}^2 \hat{M}^2 (U-\hat{c})^2]\,\Pi = 0,
\end{eqnarray}
where $\hat{c}=\hat{\om}/\hat{\al}$\/, and the prime indicates the derivative
with respect to $\eta$\/. Equation (\ref{comp1}) is easily turned
into an analogous two-dimensional disturbance equation by Squire transform
$$\al^2 = \hat{\al}^2 + \hat{\be}^2\,,\quad \hat{\al}=\al\cos \theta\,,\quad
\hat{\be} = \al\sin \theta\,,$$
$$\al M = \hat{\al}\hat{M}\,,\quad \hat{c}=c\,,\quad
\hat{\om} = \om \cos \theta\,, $$
where $\theta$ is the angle between the disturbance wavenumber and the mean
flow direction. Applying this transform to (\ref{comp1}) yields
\begin{equation}\label{comp2}
\Pi'' -\frac{2 U' \Pi'}{U-c}
- \al^2 T [T - M^2 (U-c)^2]\Pi = 0\,.
\end{equation}
Whatever $\;\Pi(\al,c)\;$ is a solution of (\ref{comp2}), $\;\Pi(-\al,c)\;$ and
$\:\Pi^*(\al^*,c^*)$ are also solutions, so that we can restrict the analysis
in the $(\al, c)$-plain
to $\al_r\geq 0$ and $c_i \geq 0$, where the indices $r$ and $i$ indicate the
real and imaginary parts of a quantity.

The boundary conditions for $\Pi$ are obtained by considering the limiting
forms of (\ref{comp2}) as $\eta \to \pm \infty$, that is by taking
$U'=0$ in (\ref{comp2}). The required solutions are
exponentials of the form
\begin{equation}\label{bcK}
\Pi \to \exp(\mp \Om_{\pm} \eta) \quad \mbox{at} \quad \eta \to \pm\infty\,,
\end{equation}
\begin{eqnarray*}\mbox{where}\quad\Om_+^2 &=&\al^2 [1-M^2(1-c)^2]\,,\\
      \Om_-^2 &=& \al^2 \be_T [\be_T-M^2(\be_U-c)^2]\,.
      \end{eqnarray*}
The linear theory leads to consider small-amplitude perturbations, 
which are weakly unstable oscillations characterizing so\-nic waves escaping
from the shear layer, in which $\Re(v_{gy}) >0$ when $y >0$ and
$\Re(v_{gy})<0$ when $y <0$; $v_{gy}$ is the $y$\/-component of the
group velocity. This warrants an exponential decrease of the wave energy
away from the shear layer. The non-radiation condition for unstable
oscillations comes down to $\Re(\Om_+) >0$ and $\Re(\Om_-) >0$, where $\Re$
indicates the real part of a quantity.
A neutral mode with a subsonic convective Mach number decays exponentially
with $\eta$ but radiates outward with a supersonic convective Mach number
like a sound wave. An unstable supersonic mode may exponentially decay
but more slowly than the subsonic modes.

As the profiles $(U,T)$ have constant ``tails,'' the un\-bo\-un\-ded-flow
problem can come down to a finite-boundary problem in the $\eta$-range
$[\eta_-,\eta_+]$ such as $|U - U(\pm\infty)| \ll 1$ as
$|\eta|\geq \mathrm{min}[-\eta_-,\eta_+]$ (Keller's boundaries).
Conditions (\ref{bcK}) therefore become
{\setlength{\mathindent}{0pt}
\begin{equation}\label{cond1}
\Big (\frac{d}{d\eta}+\Om_+\Big )\Pi(\eta_+) = 0\,,\quad
\Big (\frac{d}{d\eta}-\Om_-\Big )\Pi(\eta_-) = 0\,.
\end{equation}}
\vspace*{-0.5cm}\section{Spectral Method}
\label{sec:spectr}
When $c$ is given, equation (\ref{comp2}) together with the boundary
conditions (\ref{cond1}) constitute an eigenvalue problem, $\alpha$ being the
spectral parameter. The solution of this problem gives $\alpha$ as a function
of $c$\/, or, assuming that this function can be inverted, $c$ and,
consequently, $\omega$ as a function of $\alpha$\/. To solve this problem we
use the following numerical method. First we introduce a new cross-stream
variable $Z$ such as
\begin{equation}
\frac{\eta}{\eta_{+}}= \mu Z + (1-\mu) Z^3,\qquad -1\leq Z \leq 1
\end{equation}
where $\eta_- = - \eta_+$ and $0 <\mu <1$. Equation (\ref{comp2}) is
rewritten as
\begin{eqnarray}
\frac{d^2\Pi}{dZ^2} &-& 2\left[ 3 (1-\mu)\frac ZE + \frac{dU/dZ}{U-c}\right]
   \frac{d\Pi}{dZ} \nonumber\\
&-& (\al\eta_+ E)^2 T[T - M^2(U-c)^2]\Pi = 0,
\end{eqnarray}
\begin{equation}
\begin{array}{cc}
\mbox{where}\hspace{0.1cm}& E = \mu + 3 (1-\mu)Z^2\,, \\
&\displaystyle \frac{dU}{dZ} = \mbox{$\frac12$}(1-\be_U)\eta_
   + E\,[1-\tanh(\eta)^2]\,.
\end{array}
\end{equation}
The boundary conditions (\ref{cond1}) take the form
\begin{equation}\label{bondc}
\frac{d\Pi}{dZ}\pm \al\,\eta_+ \,E \, \varpi_{\pm}\,\Pi = 0
   \quad \mbox{at} \;\; Z = \pm 1,
\end{equation}
where $\,\varpi_{\pm} = \Om_{\pm}/\al$\/.\, The problem comes down to the
search for the eigenvalues $\lam=\al\, \eta_+\, E(1)$ ($\Re(\lam)\geq 0$) and
eigenfunctions $\Pi$, the wave speed $c$ being given.  An approximate solution
is obtained by expanding $\Pi$ in a finite Chebyshev series. The Chebyshev
collocation method is used to discretize the problem. This method was tested
by a direct integration of (\ref{comp2}) using the fourth-order Runge-Kutta
scheme in the same way as in Blumen (1970).

\section{Convective and Absolute Instabilities}
\label{sec:convect}

The general evolution of an initial perturbation with time can be formally
expressed by a Laplace-Fourier integral
\begin{equation}
q(x,y,t) = \int_{i\sig -\infty}^{i\sig +\infty}\hspace{-7mm} e^{-i\om t}\,d\om
\int_{-\infty}^\infty\hspace{-1mm} \frac{T(\al,\om,y)}{D(\al,\om)}\,
   e^{i\al x}\,d\al ,
\end{equation}                                                              
where $\sig$ is defined to be larger than the largest growth rate
of any mode for real $\al$\/, and $q$ is the perturbation of any quantity.
The dispersion relation $\;D(\al,\om)\;$ and the function
$T(\al,\om,y)\,$ are analytic functions of $\,\al\,$ and $\,\om$\/.
The function $T(\al,\om,y)$ depends on the initial perturbation. It is not
important for what follows because the asymptotic behaviour of
$\:q(x,y,t)\:$ as $\:t \to \infty\:$ is completely determined by
$D(\al,\om)$. Note that the initial-value problem is well-posed because the
growth rate of unstable modes in an inviscid compressible shear flow is
bounded by (Chimonas 1970)
\begin{equation}
\om_i \leq \f12 |U'|_{\mathrm{max}}\,.
\label{bound}
\end{equation}
To study the asymptotics of $\;q(x,y,t)\;$, we used the Briggs' method
(Briggs 1964). In accordance with this me\-thod, this asymptotics is closely
related to the double roots of $D(\al,\om)$ considered as a function of
$\al$\/. The double $\al$-roots satisfy the system of equations
\begin{equation}
D(\al,\om) = 0, \quad \frac{\partial D}{\partial\al}(\al,\om) = 0\,.
\label{system}
\end{equation}
Due to the restriction on the paper length, we do not describe the details
of solutions $(\al,\om)$ of this system of equations. They will be given 
elsewhere. In this
paper, we will only outline the main results that we obtained.

\begin{figure}
\vspace*{-4mm}
\includegraphics[width=80mm,height=60mm]{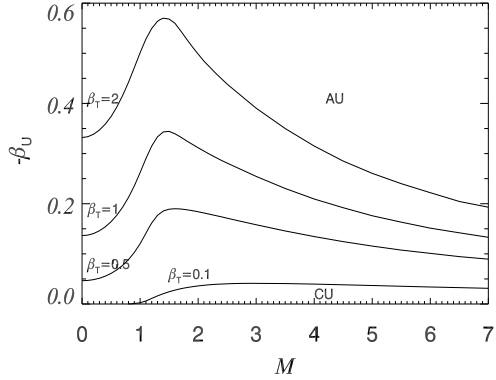}
\vspace*{-3mm}
\caption{The boundaries between the regions of parameters corresponding to the
absolute and convective instabilities in the $(M,\beta_U)$-plane for different
values of $\be_T$\/. `AU' stands for the absolute instability, and `CU' for
convective instability.}
\label{label1}
\end{figure}

\begin{figure}
\includegraphics[width=80mm,height=60mm]{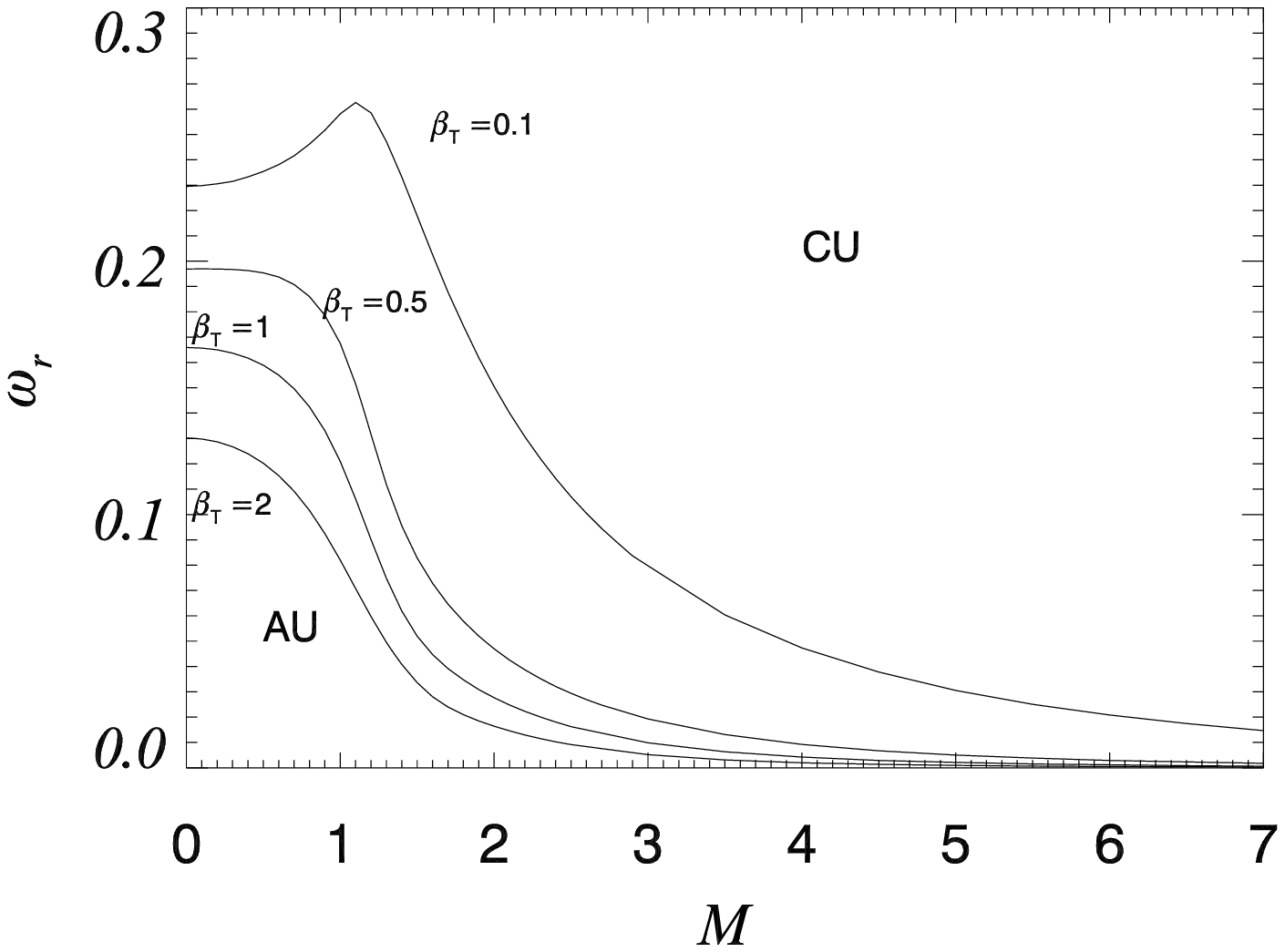}
\caption{Absolute/convective instabilities transition
in the $(\om_r, M)$ plane for various values of $\be_T$.}
\label{label2}
\end{figure}

\begin{figure}
\includegraphics[width=80mm,height=60mm]{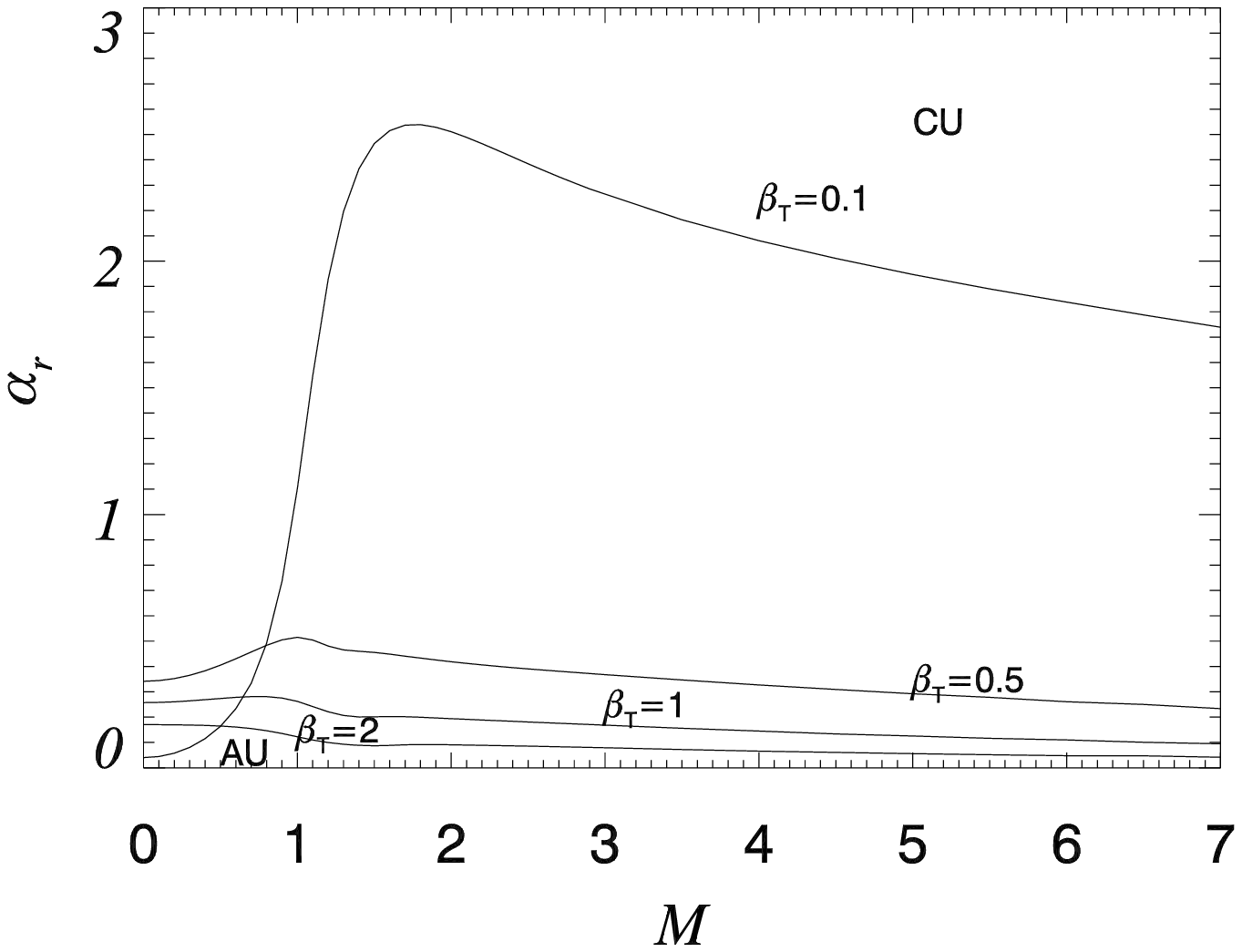}
\caption{Absolute/convective instabilities transition in the
$(\al_r, M)$ plane for various values of $\be_T$.}
\label{label3}
\end{figure}

Our main results are presented now. The figure \ref{label1} shows the
boundaries between the regions of parameters corresponding to the absolute and
convective instabilities in the $(M,\beta_U)$-plane for different values of
$\be_T$\/. A general trend is apparent whatever $\be_T$\/. There exists a
critical Mach number $M_{\rm cr}$ whose value depends on $\be_T$\/ such as when
$M < M_{\rm cr}$, increasing$\hphantom{,}$ the Mach number increases the
magnitude of counterflow$\hphantom{,}$ necessary$\hphantom{,}$ to$\hphantom{,}$
cause the absolute
instability. But as $M > M_{\rm cr}$, the situation is opposite: the larger
the Mach number is,  the smaller the magnitude of counterflow is. 
Cooling the slow stream,
i.e.\ decreasing $\be_T$\/, extends the domain of the absolute instability.  
For $\be_T$ sufficiently small, the shear layer may become absolutely unstable
in the subsonic range even when the streams are co-flowing, i.e.\ $\be_U > 0$.
In particular, the thre\-shold value of $\be_U$ becomes positive for 
$\be_T=0.1$ when $M<0.8$, and reaches a value of order $10^{-3}$.
A finer accuracy and a stronger resolution of the spectral method were used
for these
very small values of $\be_U$\/. All curves tend to zero as  
$M\to \infty$\/. This implies that the flow is absolute unstable for any value
of $\be_U < 0$ if $M$ is sufficiently large.
Figure \ref{label2} shows the transition frequency as a function
of $M$ for chosen values of $\be_T$. A similar separation of the evolution
of $\om_r$ with $M$ occurs according to whether $M$ is less or greater than 
 $M_{\rm cr}$. In the first range, $\om_r$ is quasi-constant, then it decays
rapidly for $M\sim M_{\rm cr}$. When the motion is supersonic, 
the decrease is smoother. When $\be_T=0.1$ however, 
the behaviour of $\om_r$ is different, it is increasing up to $M=M_{\rm cr}$ 
and then decreasing. The hotter the slow stream
is, the smaller the transition frequency is. The asymptotic frequency for
infinite Mach number is zero.  
Figure \ref{label3} displays the evolution of the wavenumber as a function of
$M$. This does not vary substantially for $M<M_{\rm cr}$ when $\be_T$ is
of order one, but it increases steeply when $\be_T$ is small.
The wavenumber $\al_r$ increases as $\be_T$ decreases.
We expect the high instability of the cooled-slow-stream mixing layers 
to be attenuated by viscous and thermal diffusions.

\section{Application to Heliopause Stability}
\label{sec:helio}

\begin{figure}
\begin{center}
\includegraphics[width=70mm,height=53mm]{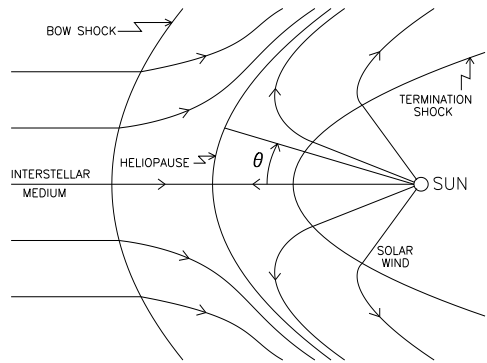}
\end{center}
\vspace*{-2mm}
\caption{Schematic of the heliosphere}
\label{helio}
\end{figure}

Collision of the solar wind with the interstellar medium results in the
interaction region consisting of two shocks and a tangential discontinuity
between them (Baranov et al. 1970, 1976, 1979; Baranov 1990). This region is
shown in Fig.~\ref{helio}. The external shock, called the bow shock,
decelerates the supersonic flow of the interstellar medium.$\hphantom{,}$
The internal shock, called the termination shock, decelerates the supersonic
flow of the solar wind. The tangential discontinuity, called the heliopause,
separates the decelerated solar wind and interstellar medium flows.

The heliopause stability was a subject of an intensive study for more than two
decades. In the early studies, the normal mode analysis combined with the local
analysis (the WKB approximation) was used (e.g., Fahr et al. 1986; Baranov et
al. 1992; Chalov, S.V. 1996). Recently, Ruderman et al. (2004) made an attempt
to study the absolute and convective instability of the heliopause. However,
since these authors used the approximation of an incompressible fluid, they
only managed to study the stability of near flanks of the heliopause (the
region corresponding to $|\theta| \lesssim 30^\circ$ in Fig.~\ref{helio}). At
the near flanks, the plasma flows at both sides of the heliopause are strongly
subsonic, so that the approximation of an incompressible fluid can be used to
describe the plasma motion. However the flow Mach number monotonically
increases with the distance from the apex point, and eventually the flow
becomes supersonic at far flanks, which implies that the account of
compressibility is of crucial importance. An immediate implication of the
results presented in the previous section is that the whole heliopause is only
convectively unstable, at least with respect to short-wave\-length perturbations
when the local analysis is valid. Really, even when the ratio of temperatures
at the two sides of the heliopause, $\beta_T$\/, is very small
(which is usually the case), the instability is absolute only if
$\beta_U \lesssim 10^{-3}$\/.
However, in accordance with all numerical models of the solar
wind--inter\-stellar medium interaction $\beta_U \gg 10^{-3}$
(e.g.\ Baranov et al. 1979).    

\section{Summary and Conclusions}
\label{sec:concl}
 
In this paper, we have studied the absolute and convective instabilities of a
compressible shear flow with the hy\-per\-bolic-tangent velocity and
temperature profiles. The stability properties of the flow are determined by
three dimensionless parameters: the Mach number $M=\hat{M} \cos \theta$
calculated using the velocity and sound speed of the fast stream, and the
ratios of the velocities $\beta_U$\/, and
temperatures $\beta_T$\/, of the fast and slow streams. We plotted the
boundaries between the regions of parameters corresponding to the absolute and
convective instabilities for different values of $\beta_T$\/.

Our main results are the following. Each value of $\beta_T$ imposes a
threshold value of the Mach number $M_{\rm cr}$\/.
When $M < M_{\rm cr}$, increasing
the Mach number increases the magnitude of counterflow (i.e.\ increases the
value of $-\beta_U$) necessary to cause the absolute instability. This result
was previously obtained by Pavithran \& Redekopp (1989) and Jackson \& Grosch
(1990) for subsonic flows. When $M > M_{\rm cr}$, increasing the Mach number
decreases the magnitude of co\-unterflow necessary to cause the absolute
instability. Co\-oling the slow stream, i.e.\ decreasing $\be_T$\/, extends the
domain of the absolute instability. For very small values of $\be_T$ the
absolute instability is even possible when the streams are co-flowing, i.e.\
$\be_U > 0$, however only for very small values of $\be_U$\/,
$\be_U \lesssim 10^{-3}$\/. The magnitude of the counterflow necessary to
cause the absolute instability tends to zero ($-\beta_U \to 0$) when
$M \to \infty$\/.

We discussed the implication of the results obtained in this paper on the
stability of the heliopause, which is the surface separating the flows of the
interstellar medium and the solar wind plasma in the solar wind--interstellar
medium interaction region. Our conclusion is that the heliopause is only
convectively unstable, at least with respect to short-wavelength perturbations.
    

\end{document}